# Synthesis, crystal structure and physical properties of quasi-one-dimensional $A$Cr$_3$As$_3$ ($A$ = Rb, Cs)

Zhang-Tu Tang[1,3], Jin-Ke Bao[1,3], Liu Yi[1,3], Hua Bai[1,3], Hao Jiang[1,3], Hui-Fei Zhai[1,3], Chun-Mu Feng[1], Zhu-An Xu[1,2,3] and Guang-Han Cao[1,2,3]*

## Abstract

Recently, new Cr-based superconductors, $A_2$Cr$_3$As$_3$ ($A$ = K, Rb, Cs), have gained a strong interest because of their one-dimensional crystal structures and electron correlations. Here we report the crystal structure and physical properties of two related materials $A$Cr$_3$As$_3$ ($A$ = Rb, Cs) which are synthesized via a soft-chemical $A^+$ deintercalation in $A_2$Cr$_3$As$_3$. The new compounds remain one-dimensional (Cr$_3$As$_3$)$_\infty$ linear chains, and the interchain distance can be tuned by the incorporation of the alkali-metal cations with different sizes. The physical-property measurements indicate a local-moment behavior at high temperatures, and the moments freeze into a cluster spin-glass state below 5~6 K. No superconductivity was observed in both materials. We also found that, with increasing the interchain distance, the Cr effective moments increase monotonically, accompanied with the enhancement of semi-conductivity. Our results shed light on the understanding of occurrence of superconductivity in $A_2$Cr$_3$As$_3$.

[1] Department of Physics, Zhejiang University, Hangzhou 310027, China
[2] Stat Key Lab of Silicon Materials, Zhejiang University, Hangzhou 310027, China
[3] Collaborative Innovation Centre of Advanced Microstructures, Nanjing 210093, China
* Corresponding author (email: ghcao@zju.edu.cn)

## INTRODUCTION

The recent discovery of superconductivity in quasi-one dimensional compounds $A_2$Cr$_3$As$_3$ ($A$ = K, Rb, Cs) [1-3] has aroused immediate research interest primarily because of the significant electron correlations as well as the quasi-one-dimensional crystal structure. $A_2$Cr$_3$As$_3$ crystallizes in a hexagonal lattice, comprising of $[($Cr$_3$As$_3)^{2-}]_\infty$ double-walled subnano-tubes that are separated by columns of the $A^+$ counterions. Both experimental investigations [1-9] and theoretical calculations/analyses [10-14] indicate novel properties in the new superconductors.

The $A_2$Cr$_3$As$_3$ superconductors are very air sensitive, probably due to the existence of "crowded" $A$1 atoms in the crystal structure [1]. By topologically removing one K in K$_2$Cr$_3$As$_3$, we very recently succeeded in obtaining an air-stable compound KCr$_3$As$_3$ [15]. This new material crystallizes in a different space group of P6$_3$/m (No. 176), yet containing the similar (Cr$_3$As$_3$)$_\infty$ linear chains. Thus it is regarded as a "cousin" of K$_2$Cr$_3$As$_3$. The physical property measurements indicate that KCr$_3$As$_3$ does not show superconductivity, but exhibits a cluster spin-glass state instead.

Here we report the synthesis and characterizations of two new members in the $A$Cr$_3$As$_3$ series with $A$ = Rb and Cs. Our measurements show that all the "133" compounds, including previously synthesized KCr$_3$As$_3$, have similar physical properties with a cluster spin-glass ground state, just like that the "233" family all superconduct at low temperatures. Nevertheless, some variations of some physical quantities, such as the resistivity value and magnitude of the local moments, are also evident, which suggests the importance of dimensionality (or interchain coupling). Our results supply helpful references, which may shed light on understanding of the superconductivity in $A_2$Cr$_3$As$_3$.

## EXPERIMENTAL

Polycrystalline samples of $A$Cr$_3$As$_3$ were synthesized by reacting $A_2$Cr$_3$As$_3$ polycrystals with water-free ethanol, similar to the previous report [15]. First, the $A_2$Cr$_3$As$_3$ polycrystalline pellets were prepared by multi-step solid-state reactions in evacuated containers [2,3]. The resultant products are single phase, as checked by powder X-ray diffraction (XRD). The as-prepared pellets were immersed into water-free ethanol at room temperature, and bubbles of hydrogen gas were found to release from the liquid. After the reaction for two days, the pellets were washed by ethanol, and then evacuated in vacuum to remove the remaining ethanol. In order to get the pure "133" phase, the process was repeated several times. We note that the obtained CsCr$_3$As$_3$ sample is much loose and friable, compared with RbCr$_3$As$_3$. By weighing a regularly-shaped sample, the density of RbCr$_3$As$_3$ (CsCr$_3$As$_3$) was measured to be 82(3)% [73(3)%] of the theoretical one based on the XRD results.

Powder XRD experiments were carried out at room temperature on a PANalytical x-ray diffractometer (Empyrean Series 2) with a monochromatic CuK$\alpha_1$ radiation. The electrical resistivity was measured using standard four-terminal method. The polycrystalline samples were cut into a thin rectangular bar, and silver paste was used to attach gold wires onto the samples' surface. The magnetizations were measured on a Quantum Design Magnetic Property Measurement System (MPMS-5). The specific-heat capacity was measured by a relaxation technique on a Quantum Design Physical Properties Measurement System (PPMS-9) using a thin square-sharp sample. The heat-capacity addenda from the sample holder and grease were deducted.

## RESULTS AND DISCUSSION

Based on the previous studies [1-3,15], the two-step chemical reactions associated with the synthesis can be written as follows,

(1) $2A + 3Cr + 3As \rightarrow A_2Cr_3As_3$ (~700°C in vacuum);

(2) $A_2Cr_3As_3 + C_2H_5OH \rightarrow ACr_3As_3 + AOC_2H_5 + 1/2H_2\uparrow$.

The second reaction keeps the identical $(Cr_3As_3)_\infty$ linear chains (see below). Therefore, the reaction is actually a deintercalation process, which is in general a topotactic transformation without breaking the main framework of the crystal structure. Thus it is expected to produce single crystals using the similar route in the future.

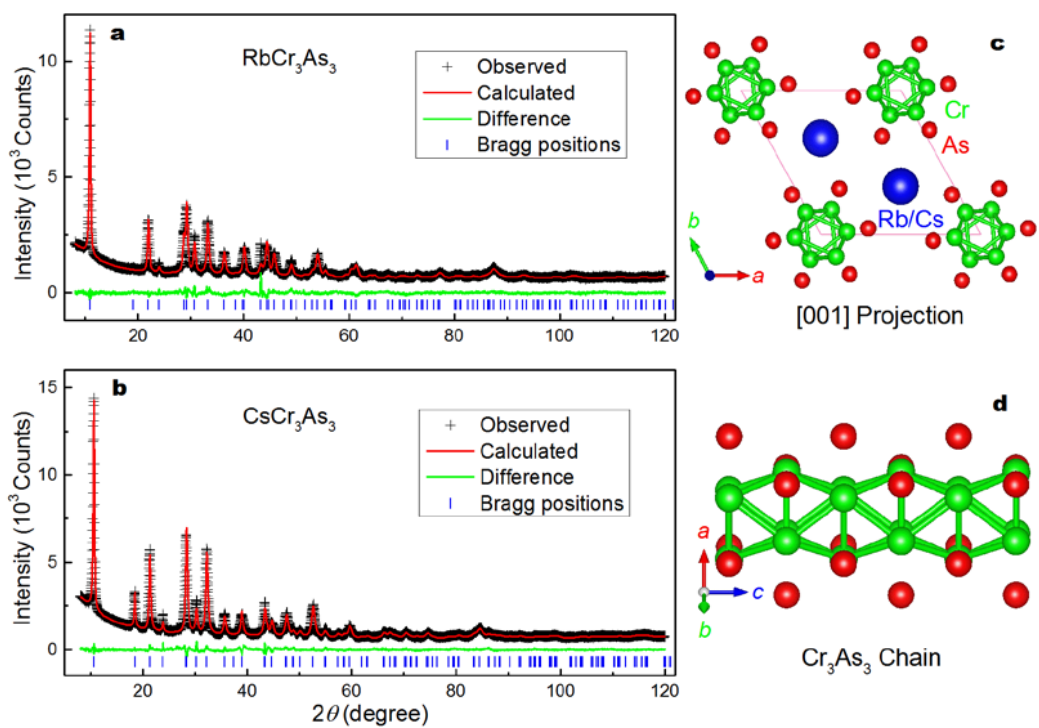

**Figure 1** Room-temperature powder X-ray diffraction patterns and their Rietveld refinement profiles for $RbCr_3As_3$ (a) and $CsCr_3As_3$ (b). Panels (c) and (d) display the crystal structure from two different perspectives, highlighting the one-dimensional $(Cr_3As_3)_\infty$ double-walled subnano-tubes.

Table 1 Experimental crystallographic parameters refined by a Rietveld analysis of the X-ray diffraction data collected at room temperature for $ACr_3As_3$. The atomic coordinates are ($x, y, 0.25$).

| | RbCr₃As₃ | CsCr₃As₃ | | | | |
|---|---|---|---|---|---|---|
| **Space group** | $P6_3/m$ (#176) | $P6_3/m$ (#176) | As (6h) | $x$ | 0.3342(4) | 0.3182(3) |
| | | | | $y$ | 0.0542(3) | 0.0425(2) |
| $a$ (Å) | 9.334 (2) | 9.632 (1) | Cr (6h) | $x$ | 0.1522(5) | 0.1521(3) |
| $c$ (Å) | 4.1775 (7) | 4.1814 (6) | | $y$ | 0.1816(4) | 0.1711(2) |
| **Coordinates** | | | A (2c) | | $x$=1/3 | $y$=2/3 |

| Cr–Cr Bond distances (Å) | | |
|---|---|---|
| Inter-plane Cr–Cr | 2.616(3) | 2.611(1) |
| In-plane Cr–Cr | 2.729(5) | 2.710(3) |

Fig. 1 shows the room-temperature XRD patterns for the $A$Cr$_3$As$_3$ powder samples. Both can be well indexed by a hexagonal lattice with unit-cell parameters close to those of KCr$_3$As$_3$ [15], and no obvious impurity phase can be identified. Using the same structure (space group $P6_3/m$) as the starting point, we successfully made the Rietveld structural refinements [16], with a reliable factor of $R_{wp}$=4.9%, a goodness of fit $\chi^2$=2.3 for RbCr$_3$As$_3$ and $R_{wp}$=4.4%, $\chi^2$=2.3 for CsCr$_3$As$_3$, respectively. The resulting crystallographic data are listed in Table 1.

The crystal structure of the $A$Cr$_3$As$_3$ system is shown in Fig. 1c and 1d. The quasi-one-dimensional (Cr$_3$As$_3$)$_\infty$ double-walled subnano-tubes are kept, although half of the Rb$^+$ or Cs$^+$ was deintercalated. The linear chains comprise face-sharing Cr$_6$ octahedra, which are surrounded by As$^{3-}$ ions. There is only one equivalent site for the $A^+$ cation. Each $A^+$ cation is coordinated to nine As$^{3-}$ anions in a tricapped triangular prisms. Notably, compared with the $A_2$Cr$_3$As$_3$ superconductors, the (Cr$_3$As$_3$)$_\infty$ chains rotate a small angle along the $c$ axis, which leads to a change in space group and point group. Such rotations could lead to significant changes in band structures, as shown by a very recent calculation for KCr$_3$As$_3$ [17].

Fig. 2 shows the change in crystal structure as functions of the ionic radii [18] of the alkali-metal cations. The $a$ axis increases remarkably and almost linearly, while the $c$ axis hardly changes. This reflects that the (Cr$_3$As$_3$)$_\infty$ chains basically remains unchanged, and the interchain distance (the same as the $a$ value) can be modified by the alkali-metal elements. Indeed, the Cr-Cr bond distances have only a slight change, as shown in Fig. 2b.

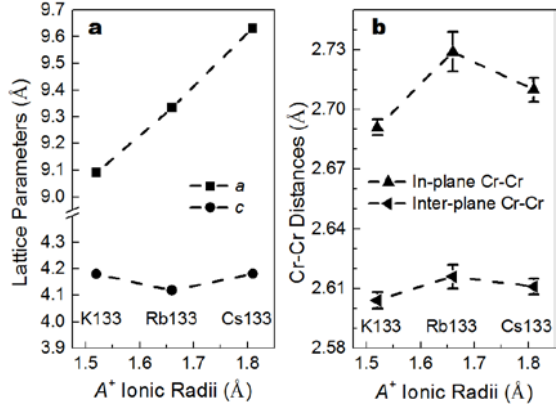

**Figure 2** Influence of the size of the alkali-metal element on the crystal structure of $A$Cr$_3$As$_3$ ($A$=K, Rb, Cs).

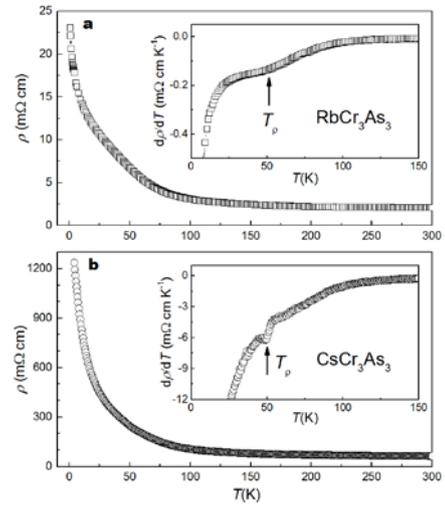

**Figure 3** Temperature dependence of electrical resistivity of the $A$Cr$_3$As$_3$ ($A$=Rb, Cs) polycrystalline samples. The insets plot the derivative of resistivity, showing an anomaly at ~50 K.

Fig. 3 shows the temperature dependence of electrical resistivity, $\rho(T)$, for the $A$Cr$_3$As$_3$ polycrystalline samples. Similar to that in KCr$_3$As$_3$ [15], the resistivity of $A$Cr$_3$As$_3$ has a weak temperature dependence above 100 K. Below 100 K, the resistivity shows a semiconducting-like behavior. However, the $\rho(T)$ data do not obey a thermally-activated relation, $\rho(T)=\rho_0\exp[-E_a/(k_BT)]$, where $E_a$ denotes the excitation energy. Neither do they follow the variable-range hopping formula originally proposed by Mott [19]. The room-temperature resistivity of RbCr$_3$As$_3$ is ~2 mΩ cm, comparable to that of KCr$_3$As$_3$ (~1 mΩ cm). CsCr$_3$As$_3$ shows a much higher room-temperature resistivity (~60 mΩ cm), which seems to be related to the sample's looseness. Nevertheless, the resistivity ratio, $\rho_{4K}/\rho_{300K}$, also has the largest value in CsCr$_3$As$_3$. The remarkable difference in $\rho(T)$ does not seem to be ascribed to a grain boundary effect.

It is expect that the intrinsic electrical transport is highly anisotropic, given the quasi-one-dimensional crystal structure. Namely, the conductivity along the chain direction would be larger, and the conductivity perpendicular to the $c$ axis would be relatively small. Assuming metallic conductions along the chains (because of non-zero density of state at the Fermi level evidenced by the related band-structure calculations [17] and specific-heat measurements below), the measured high resistivity as well as the semiconducting-like temperature dependence could represent the interchain transport property. Another possibility of the pronounced semiconductivity may be associated with an enhanced Anderson localization due to the weakened interchain coupling. Note that the d$\rho$/d$T$ data show a dip at $T_\rho \sim 50$ K, which is slightly lower than that of KCr$_3$As$_3$ [15]. This anomaly is probably related to the formation of Cr-spin clusters [15].

Fig. 4a shows the temperature dependence of dc magnetic susceptibility ($\chi=M/H$) under an external field of 1 kOe for the bulk sample of RbCr$_3$As$_3$. The high-temperature $\chi(T)$ curve exhibits a Curie-Weiss paramagnetic behavior. We thus fit the data (100 K– 280 K) by an extended Curie-Weiss formula, $\chi=\chi_0+C/(T-\theta_\mathrm{p})$, where $\chi_0$ is a temperature-independent term, $C$ represents the Curie constant, and $\theta_\mathrm{p}$ is the paramagnetic Curie-Weiss temperature. The fitted $C$ value gives an effective magnetic moment of 1.22 $\mu_\mathrm{B}$/Cr, twice as large as that of KCr$_3$As$_3$ (0.68 $\mu_\mathrm{B}$/Cr ) [15]. Besides, the fitted $\theta_\mathrm{p}$ value is −78.8 K, implying dominantly antiferromagnetic interactions between the Cr local moments.

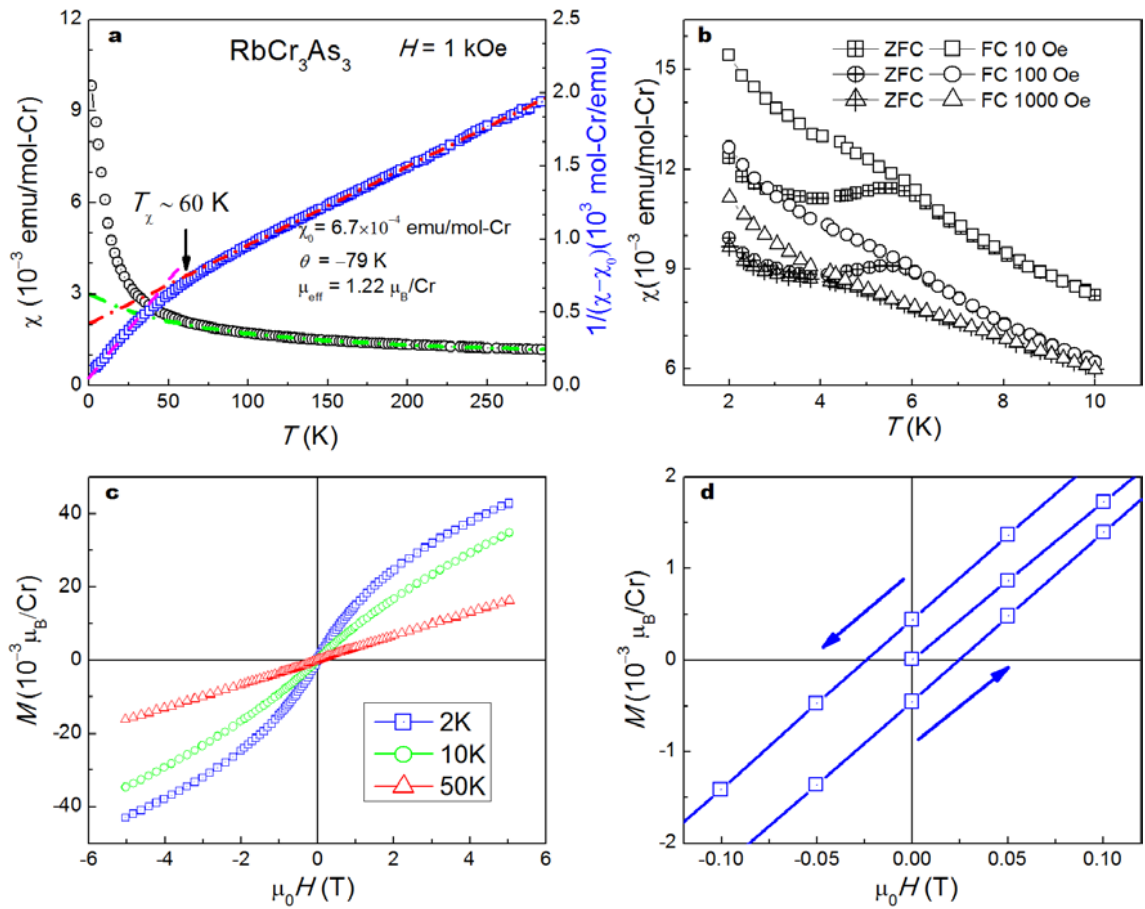

**Figure 4** (a) Temperature dependence of magnetic susceptibility for RbCr$_3$As$_3$. The dashed lines represent the Curie-Weiss fit. (b) Low-temperature magnetic susceptibility with zero-field-cooled (ZFC) and field-cooled (FC) procedures under different external fields. (c) Isothermal magnetizations at different temperatures. (d) An enlarged plot of field dependence of magnetization at 2 K.

The data of $1/(\chi-\chi_0)$ obviously deviates from linearity below $T_\chi \sim 60$ K, below which a broad transition appears, corresponding to the broad anomaly in $d\rho/dT$ shown in Fig. 3a. The Curie-Weiss fit for low-temperature (8 K – 50 K) data yields a reduced effective moment of 0.67 $\mu_B$/Cr, which suggests formation of spin clusters. If the spin clusters are composed of six nearby Cr atoms (a $Cr_6$ octahedron), the effective moments become 1.64 $\mu_B/Cr_6$, equivalent to a localized $S=1/2$ spin. Additionally, the paramagnetic Curie-Weiss temperature changes into 1.5 K, indicating a weak ferromagnetic interaction between the spin clusters.

At lower temperatures, the spin clusters freeze, as indicated in Fig. 4b-4d. The $\chi(T)$ data at low magnetic fields show an obvious bifurcation at $T_f = 6$ K between the zero-field cooling (ZFC) and field cooling (FC) data. With increasing field, the divergence tends to diminish, and $T_f$ shifts to lower temperatures. The isothermal magnetization, $M(H)$, shows non-linear relation at low temperatures. An obvious magnetic hysteresis can be seen below $T_f$, in accordance with the spin freezing. Meanwhile, the magnetization does not saturate even under a magnetic field of 5 T, where the magnetization is less than 0.05 $\mu_B$/Cr. This suggests absence of long-range ferromagnetic ordering.

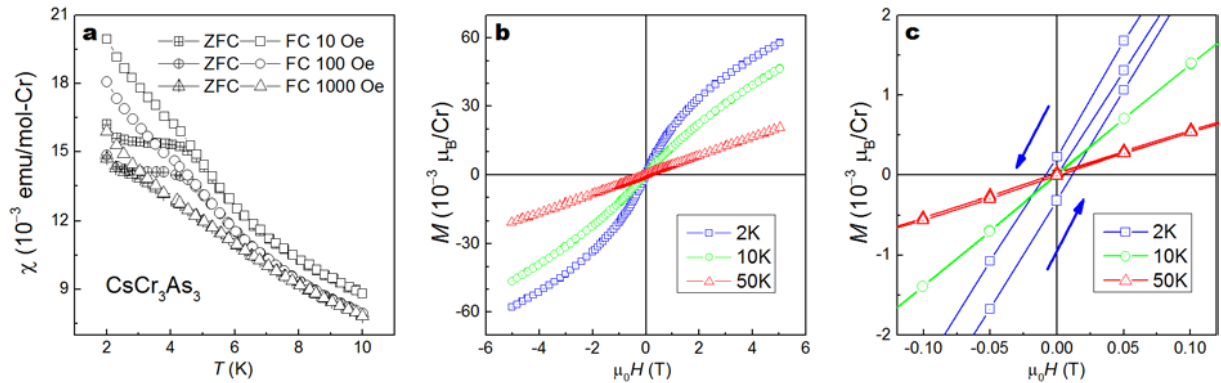

**Figure 5** (a) Low-temperature magnetic susceptibility under different external fields for $CsCr_3As_3$. (b) Isothermal magnetizations at different temperatures. (c) An enlarged plot of field dependence of magnetization, showing the magnetic hysteresis.

We also measured the dc magnetic susceptibility and isothermal magnetization for $CsCr_3As_3$, as shown in Fig. 5. Very similar behavior was observed, which points to the spin cluster freezing below 5 K. The high-temperature susceptibility (not shown here) indicates a larger effective moment of 1.6 $\mu_B$/Cr, which is consistent with the weakest interchain coupling in $CsCr_3As_3$.

Fig. 6 shows the temperature dependence of specific heat data of $RbCr_3As_3$ and $CsCr_3As_3$, plotted in $C/T$ vs $T^2$. No specific-heat jump associated with a superconducting transition can be seen, consistent with the above electrical and magnetic measurements. The prominent feature is the broad round-shape $C/T$ vs $T^2$ relation, instead of an essentially linear region in their superconducting "cousins" $A_2Cr_3As_3$ [2,3]. As argued in our previous report [15], this anomaly probably results from the cluster spin-glass state described above.

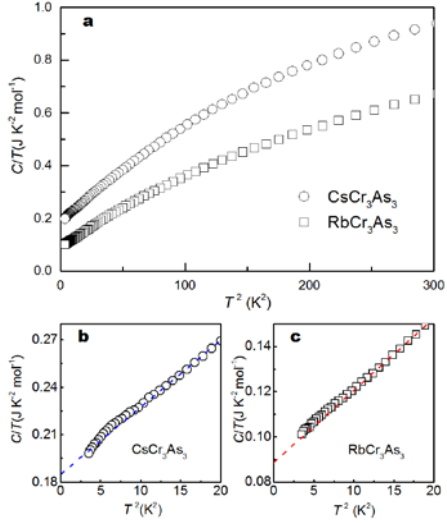

**Figure 6** Low-temperature specific-heat data for RbCr$_3$As$_3$ and CsCr$_3$As$_3$, plotted with $C/T$ vs $T^2$.

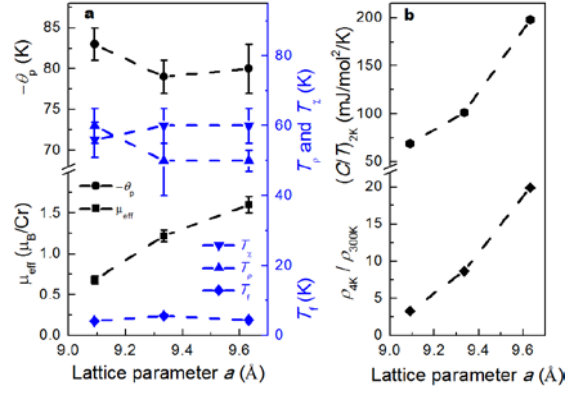

**Figure 7** Extracted physical parameters as functions of the unit-cell constants $a$ for $A$Cr$_3$As$_3$ ($A$=K, Rb, Cs).

If we make a linear fit for the data below 4 K, a zero-temperature specific-heat coefficient can be extrapolated, which is 89 mJ K$^{-2}$ mol-fu$^{-1}$ and 185 mJ K$^{-2}$ mol-fu$^{-1}$ for RbCr$_3$As$_3$ and CsCr$_3$As$_3$, respectively. Given the existence of cluster spin glass, the extrapolated specific-heat coefficient does not represent the Sommerfeld coefficient, since the magnetic contribution $C_m$ of a spin glass is basically linear with temperature for $T < T_f$ [20]. Nevertheless, the "real" Sommerfeld coefficient would not go to zero, otherwise the magnetic entropy becomes unphysically large. This means that there is still appreciably high density of states at the Fermi level of RbCr$_3$As$_3$ or CsCr$_3$As$_3$, irrespective of their apparent semi-conductivity

Fig. 7 summarizes some physical-property parameters as functions of the lattice constants $a$ that measures the interchain distance. From $A$ = K, Rb, to Cs, the magnitude of effective magnetic moments for Cr increases significantly, nevertheless, the paramagnetic Neel temperature which reflects antiferromagnetic interactions hardly changes. The characteristic temperatures $T_f$, $T_\rho$ and $T_\chi$ are also insensitive to the interchain distance. Meanwhile, the semi-conductivity, here characterized by the resistivity ratio of $\rho_{4K}/\rho_{300K}$, also increases with the interchain distance. Notably, the $C/T$ value at 2 K shows the similar tendency. Given the same "doping level" in $A$Cr$_3$As$_3$ (the apparent Cr valence is +2.67), such changes should be mainly ascribed to the interchain coupling. In general, the reduced interchain coupling enhances spin fluctuations, Anderson localizations, and electron correlations (the related bandwidth is narrowed). Therefore, the results can be qualitatively interpreted in terms of an enhanced one dimensionality. The latest first-principles calculations for KCr$_3$As$_3$ [17] indicate an "interlayer antiferromagnetic (IAF)" ground state where the Cr spins in the planar Cr triangles align ferromagnetically to form a larger "block spin". It was found that there is a strong quasi-one-dimensional van Hove singularity near the Fermi energy. The pronounced $C/T$ value at 2 K might be related to the van Hove singularity.

**CONCLUDING REMARKS**

In summary, two new quasi-one-dimensional Cr-based compounds, $RbCr_3As_3$ and $CsCr_3As_3$, were successfully synthesized. The crystal structure remains the characteristic $(Cr_3As_3)_\infty$ linear chains. Meanwhile, the interchain distance increases appreciably owing to the large size of $Rb^+$ and $Cs^+$ cations. Like their sister compound $KCr_3As_3$, both materials do not superconduct, but show a cluster spin-glass state below 5~6 K. Importantly, the Cr localized moment increases with the expansion of interchain distance. This further confirms the existence of Cr local-moment components in the "133" system, which implies the relevance of spins to the novel superconductivity in "233" materials.

**Acknowledgements** This work was supported by the Natural Science Foundation of China (Nos. 11190023 and 11474252), the National Basic Research Program (No. 2011CBA00103), and the Fundamental Research Funds for the Central Universities of China.

**Author contributions** Cao GH and Bao JK coordinated the work. Tang ZT and Cao GH wrote the paper in discussion with other co-authors. Tang ZT performed most of the experiments, including the growth, characterizations, and physical property measurements with assistance by Bao JK, Liu Y, Bai H, Hao J, Zhai HF and Feng CM.

**Conflict of Interest** The authors declare that they have no conflict of interest.